\documentclass[fleqn,usenatbib]{mnras}

\usepackage[utf8]{inputenc}
\usepackage{float}
\usepackage{graphicx}
\DeclareUnicodeCharacter{2212}{\textminus}

\usepackage{newtxtext,newtxmath}

\usepackage[T1]{fontenc}

\DeclareRobustCommand{\VAN}[3]{#2}
\let\VANthebibliography\thebibliography
\def\thebibliography{\DeclareRobustCommand{\VAN}[3]{##3}\VANthebibliography}

\usepackage{graphicx}	
\usepackage{amsmath}	
\usepackage{float}

\title[]{The fractions of reflection effect systems detected in different subclasses of hot subdwarfs.}

\author[Ruijie He et al.]{
Ruijie He,$^{1,2}$\thanks{E-mail: heruijie@ynao.ac.cn}
Xiangcun Meng,$^{1}$\thanks{E-mail: xiangcunmeng@ynao.ac.cn}
and Zhenxin Lei$^{3,4}$
\\
$^{1}$International Centre of Supernovae (ICESUN), Yunnan Key Laboratory of Supernova Research, Yunnan Observatories, Chinese Academy of Sciences \\(CAS), Kunming 650216, China\\
$^{2}$University of Chinese Academy of Sciences, Beijing 100049, PR China\\
$^{3}$ Key Laboratory of Stars and Interstellar Medium, Xiangtan University, Xiangtan 411105, PR China\\
$^{4}$ Physics Department, Xiangtan University, Xiangtan 411105, PR China
}

\date{Accepted 2025 September 15. Received 2025 September 15; in original form 2025 April 22}

\pubyear{2025}

\begin{document}
\label{firstpage}
\pagerange{\pageref{firstpage}--\pageref{lastpage}}
\maketitle

\begin{abstract}

Hot subdwarfs with close low-mass M-type or brown dwarf companions usually show the reflection effect and often also eclipses. Through inspecting the light variations, we estimate the fractions of reflection effect systems detected in different subclasses of hot subdwarfs and constrain their possible formation channels. We noticed that none of the helium-rich hot subdwarfs were found with the reflection effect. Most of them might form through the merger channel. About 13\% of the He-poor hot subdwarfs on the extreme horizontal branch (EHB) with $T_{\rm eff}$ $\sim$ 25,000 -- 33,000 K are discovered with the reflection effect. Nevertheless, the cooler hot subdwarfs located on the top of the EHB, those grouped on the bottom of the EHB, and those located above the EHB with $T_{\rm eff}$ $\sim$ 45,000 -- 75,000 K exhibit lower fractions of reflection effect systems of about 2\%. These three subclasses of hot subdwarfs might contain higher fractions of single stars. Hot subdwarfs below the EHB also show a low fraction of reflection effect systems. They might have a higher fraction of stars with close WD companions. A similar fraction of reflection effect systems to those on the EHB with $T_{\rm eff}$ $\sim$ 25,000 -- 33,000 K is found in hot subdwarfs located above the EHB with $T_{\rm eff}$ $\sim$ 35,000 -- 45,000 K. We also discovered that hot subdwarfs close to the Galactic Plane might have a higher fraction of stars with the reflection effect than those at higher latitudes.

\end{abstract}




\section{Introduction}

Hot subdwarfs are generally thought to be in the helium-core (He-core) or He-shell-burning stage. They reside on the extreme blue end of the horizontal branch (HB) in the Hertzsprung-Russell (HR) diagram, which are also known as the extreme horizontal branch (EHB) stars in some Galactic globular clusters (GGCs). Canonical hot subdwarfs have masses of about 0.5 $M_\odot$ and are composed of a helium-core and extremely thin hydrogen envelopes \citep[< 0.02 $M_\odot$,][]{Heber.etal.1986}. Their effective temperatures ($T_{\rm eff}$) range between about 20,000 -- 80,000 K \citep[see][for a detailed review]{Heber.etal.2009,Heber.etal.2016}. According to their atmospheric chemical compositions, they can be classified as He-poor hot subdwarfs ($\log n{\rm (He)}/n{\rm (H)} < -1$), intermediate He-rich hot subdwarfs ($ \leq -1 \log n{\rm (He)}/n{\rm (H)} \leq 0.6$), and extremely He-rich hot subdwarfs ($ \log n{\rm (He)}/n{\rm (H)} > 0.6$). About one-third to two-thirds of He-poor hot subdwarfs are believed to be in short-period binaries \citep{Maxted.etal.2001,Napiwotzki.etal.2004,Copperwheat.etal.2011,Kawka.etal.2015,Geier.etal.2022,Ruijie.etal.2024}. But only a few percent of He-rich hot subdwarfs exhibit significant radial velocity (RV) variations \citep{Napiwotzki.etal.2004,Geier.etal.2022,Ruijie.etal.2024}.

Hot subdwarfs play very important roles in many fields in astrophysics. They may be the main origin of the ultraviolet upturn in elliptical galaxies \citep{O'Connell.etal.1999,Han.etal.2007}. Pulsating hot subdwarfs are ideal objects for asteroseismology to study their interior structures and can help us to understand their evolutionary history \citep{Charpinet.etal.2011,Baran.etal.2012,Zong.etal.2018}. Short-period binary hot subdwarfs that contain a massive white dwarf (WD) are good candidates for the progenitors of type Ia supernovae (SNe Ia) \citep{Maxted.etal.2000,Geier.etal.2007,Geier.etal.2013,Wang.etal.2010,Pelisoli.etal.2021}. On the other hand, the fast single hot subdwarfs might be surviving companions of SNe Ia from the WD+MS channel \citep[e.g.,][]{Meng.etal.2019,Meng.etal.2021,Geier.etal.2015,Geier.etal.2024}. Short-period binary hot subdwarfs also play an important role in constraining the binary evolution and the CE ejection process \citep{Ivanova.etal.2013,Ge.etal.2022,Ge.etal.2024,Kramer.etal.2020}, as well as in providing reliable gravitational wave sources for future researches \citep{Kupfer.etal.2018,Kupfer.etal.2024,Lin.etal.2024}.

The companions of short-period binary hot subdwarfs are mainly M-type main-sequence (dM), WD, or brown dwarf (BD) stars \citep[e.g.,][]{Kupfer.etal.2015,Heber.etal.2016,Schaffenroth.etal.2022,Schaffenroth.etal.2023}. \citet{Schaffenroth.etal.2022} found that the periods of most hot subdwarfs with close low-mass M-type or BD companions are distributed between 2 and 20 h, and the period distribution has a broad peak from 2 to 8 h and falls off quickly at about 8 h, while hot subdwarfs with WD companions show a broad period distribution from about 1 hour to 27 days. Another distinct type of hot subdwarfs are long-period composite objects. Because these stars usually have F-, G-, or K-type companions, an infrared (IR) excess can be detected in their spectral energy distributions \citep{Vos.etal.2020,Heber.etal.2018,Solano.etal.2022}. 

Most of the hydrogen envelopes of hot subdwarfs' progenitors need to be stripped before the He-core-burning phase, which is very difficult to achieve in the context of single-star evolution. Therefore, binary evolution scenarios are usually proposed to explain their formation. Short-period binary hot subdwarfs are believed to form through the common-envelope (CE) ejection channel \citep{Han.etal.2002,Han.etal.2003,Xiong.etal.2017,Ge.etal.2024,Ge.etal.2022}. Long-period hot subdwarf binaries are generated from the stable Roche-lobe overflow (RLOF) channel \citep{Han.etal.2002,Han.etal.2003,Vos.etal.2020,Chen.etal.2013,Gotberg.etal.2018}. Single hot subdwarfs are traditionally considered to be the product of merger channels in short-period binaries, such as the double He-WDs merger \citep{Webbink.etal.1984,Zhang.etal.2012,Hall.etal.2016,Schwab.etal.2018}, the merger of a red giant star
(RGB) or a He-WD with its dM companion \citep{Politano.etal.2008,Zhang.etal.2017}.

Different types of hot subdwarfs can exhibit different characteristics of light variations. Hot subdwarfs with close low-mass M-type or BD companions can show obvious quasi-sinusoidal variability in their light curves owing to the reflection effect. These light variations usually exhibit broad minima and sharper maxima with amplitudes from a few percent up to about 20\% \citep[e.g.,][]{Schaffenroth.etal.2022,Schaffenroth.etal.2023,Barlow.etal.2024}. Eclipsing reflection effect systems are also called as HW Vir systems. Compact hot subdwarf binaries with WD companions can exhibit light variations due to the ellipsoidal deformation or Doppler beaming. They usually show tiny light variations (about 0.1\% to a few percent), and the depths of the two minima and/or the two maxima are usually different \citep{Kupfer.etal.2020b,Kupfer.etal.2020a,Kupfer.etal.2022,Bloemen.etal.2011,Yang.etal.2025,Pelisoli.etal.2021}. Composite hot subdwarfs usually show small amplitude of light variations, originating from spots on the surface of the cool companions \citep{Pelisoli.etal.2020}. Many hot subdwarfs also show variability due to the pulsations. They are mainly composed of two types of pulsators. The rapidly pulsating hot subdwarfs are called as pressure mode pulsators with typical periods of a few minutes. Their light variation amplitudes can reach up to a few percent \citep{Ostensen.etal.2010,Baran.etal.2023,Uzundag.etal.2024}. The slowly pulsating hot subdwarfs are named as gravity mode pulsators with typical periods of about 1 -- 2 h, and their light variations are usually below about 0.1\% \citep{Green.etal.2003,Reed.etal.2011,Uzundag.etal.2024}.

Since hot subdwarfs with close low-mass M-type or BD companions usually exhibit the most obvious light variations, they are easily and widely detected through inspecting their light curves \citep[e.g.,][]{Schaffenroth.etal.2019,Schaffenroth.etal.2022,Sahoo.etal.2020,Baran.etal.2021MNRAS.503.2157B,Baran.etal.2021,Barlow.etal.2022}. Meanwhile, it is widely accepted that these systems are formed through the first CE ejection channel, and the fraction of them among hot subdwarfs can serve as an important parameter to constrain the CE ejection process, as well as the input parameters of binary population synthesis (BPS) simulations \citep{Han.etal.2002,Han.etal.2003}. Although many reflection effect systems have been found, only \citet{Schaffenroth.etal.2018} have estimated the fraction of reflection effect systems among hot subdwarfs by using 26 significant RV variable hot subdwarfs randomly observed from the Massive Unseen Companions to Hot Faint Underluminous Stars from SDSS (MUCHFUSS) project. They found about 15\% of their sample showing the reflection effect, which corresponds to a detected fraction of about 15\% for the stars with close low-mass M-type or BD companions. However, the sample of \citet{Schaffenroth.etal.2018} is small and the fraction of reflection effect systems among hot subdwarfs without pre-sample selection is still unknown, which deserves a large number of samples to explore the fraction.

Recently, \citet{Geier.etal.2022} and \citet{Ruijie.etal.2024} found that He-poor hot subdwarfs along the different regions of the EHB exhibit distinct RV-variability fractions. Specifically, hot subdwarfs cooler than about 25,000 K, the subclass of hot subdwarfs with $T_{\rm eff}$ $\sim$ 33,000 -- 40,000 K and surface gravities ($\log g$) $\sim$ 5.7 -- 6.0, as well as hot subdwarfs with $T_{\rm eff}$ $\sim$ 45,000 -- 70,000 K all exhibit lower RV-variability fractions of about 15\%. The value is obviously less than the fraction of about 35\% for hot subdwarfs on the EHB with $T_{\rm eff}$ $\sim$ 25,000 -- 33,000 K. Furthermore, hot subdwarfs with $T_{\rm eff}$ $\sim$ 35,000 – 45,000 K located above the EHB show a similar RV-variability fraction (about 30\%) to those on the EHB at about 25,000 – 33,000 K, while the largest RV-variability fraction (about 50\%) is found in hot subdwarfs below the canonical EHB and they exhibit a higher fraction of stars with large RV-amplitudes than other subclasses
of hot subdwarfs \citep{Geier.etal.2022}. These distinct RV-variability fractions for different subclasses of hot subdwarfs indicate that they might originate from distinct formation channels. Similar to the information obtained from the RV variability, the fractions of reflection effect systems detected in different subclasses of hot subdwarfs can also serve as a parameter to evaluate the role of the first CE ejection channel that plays in the formation of the different subclasses. This might confirm the previous results or provide a new perspective for us to understand the formation of different subclasses of hot subdwarfs.

The environment in which the progenitors of hot subdwarfs are located might affect their formation and some of their properties. A clear example is that hot subdwarfs in NGC 6752 and $\omega$ Centauri have a much lower binary fraction and higher fraction of He-rich stars than the field hot subdwarfs \citep{Latour.atal.2018,Moni-Bidin.etal.2006,Moni-Bidin.etal.2008,Lei.etal.2019}. Recently, by using the light curves derived through the Zwicky Transient Facility \citep[ZTF,][]{Graham.etal.2019}, \citet{Wang.etal.2023} discovered 67 HW Vir systems, 496 sinusoidal systems (reflection effect, pulsation or rotation sinusoids), 11 eclipsing systems, and 4 ellipsoidal modulated systems in the 39,800 hot subdwarf candidates compiled by \citet{Geier.etal.2019}. They found that 3.6\% of the hot subdwarf candidates are light-variable within $-25^{\circ} < b < 25^{\circ}$, while a lower variable fraction of 2.0\% was found for hot subdwarf candidates in $\vert b \vert > 25^{\circ}$. This might imply that hot subdwarf candidates in the different regions of the Galaxy have different light-variable fractions, as well as different fractions of reflection effect systems since the tiny sinusoidal variations caused by the pulsation or rotation cannot be captured in large numbers by the ground-based telescopes but need further confirmation. At the same time, it also remains unknown whether the pure hot subdwarf sample in the different regions of the Galaxy rather than candidates will show different fractions of reflection effect systems, which might reveal the evolution histories of hot subdwarfs in the different regions of the Galaxy.

In this paper, we use the light curves from the Transiting Exoplanet Survey Satellite \citep[TESS,][]{Ricker.etal.2015} to detect the reflection effect systems, and try to estimate their fractions among different subclasses of hot subdwarfs. The structure of this paper is as follows. In Section \ref{sect.2} we introduce our data reduction. Section \ref{sect.3} describes our results and speculations about the possible formation channels of different subclasses of hot subdwarfs. In Section \ref{sect.4} we explore the fractions of reflection effect systems detected among hot subdwarfs in the different regions of the Galaxy. Our conclusions are given in Section \ref{sect.5}.

\section{DATA REDUCTION}\label{sect.2}

\subsection{Sample selection}

Our initial sample was selected from the spectroscopically confirmed hot subdwarfs catalog\footnote{\url{http://cdsarc.u-strasbg.fr/viz-bin/cat/J/A+A/662/A40}} compiled by \citet{Culpan.etal.2022}, which consists of 6,616 stars. During our selection process, the hotter central stars of planetary nebulae (CSPNe) and the post-asymptotic giant branch (post-AGB) stars that were classified as O(H), O(He), PG1159, and [WR] were removed. Additionally, hot subdwarfs identified in \citet{Lei.etal.2023,Lei.etal.2020,Lei.etal.2019,Lei.etal.2018} and \citet{Luo.etal.2021} with reliable atmospheric parameter measurements were included in our sample, and these sources were cross-matched against the spectroscopically confirmed hot subdwarfs to eliminate duplicates. The final sample consists of 6,258 hot subdwarfs, and 3,195 of them have the 2-minute or the 20-second cadence light curves observed by TESS.

\subsection{Light curves analysis}

    \begin{figure}
    {
    \centering
    \includegraphics[width=0.48\textwidth]{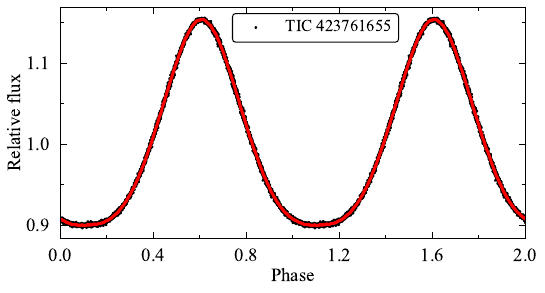}
    \includegraphics[width=0.48\textwidth]{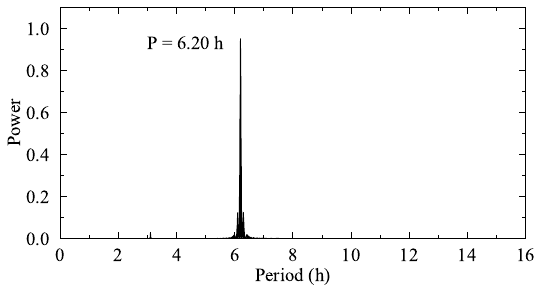}
      \caption{Sample phase-folded light curve and periodogram of a reflection effect system. Upper panel: Light curve was phase-folded to the orbital period (black points) and also binned (red points). Bottom panel: Periodogram (black lines). 
              }
         \label{Fig.1}
         }
    \end{figure}

The TESS \citep{Ricker.etal.2015} mission was first conducted in 2018. Because TESS has a wide field of view of $24^{\circ} \times 90^{\circ}$, it can obtain a large number of high-quality photometric data for many stars simultaneously. TESS observations mainly generate 30-minute, 2-minute, and 20-second cadence light curves that are available by the TESS Science Processing Operations Center (SPOC) through the Barbara A. Mikulski Archive for Space Telescopes MAST\footnote{\url{https://mast.stsci.edu/}}. Following the data reduction method adopted by \citet{Schaffenroth.etal.2022}, we used the publicly available custom script\footnote{\url{https://github.com/ipelisoli/TESS-LS}} developed by \citet{Pelisoli.etal.2020} to download the 2-minute and 20-second cadence light curves from TESS and analyze them. The pre-search data-conditioning PDCSAP\_FLUX was used, which corrected the simple aperture photometry (SAP) to remove instrumental trends, and the contributions to the aperture that are expected to come from neighboring stars other than the target of interest, given a pre-search data conditioning (PDC). In order to estimate the ratio of target flux to total flux in the TESS aperture, the CROWDSAP parameter of each star derived through the SPOC pipeline was utilized. The custom script was also used to perform sigma-clipping to exclude any measurements more than 5-sigma away from the median value. 

\begin{figure*}
    \centering
   \begin{minipage}{0.33\textwidth}
   \includegraphics[width=\textwidth]{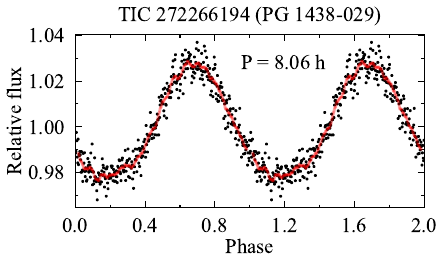}
   \end{minipage}
    \begin{minipage}{0.33\textwidth}
    \includegraphics[width=\textwidth]{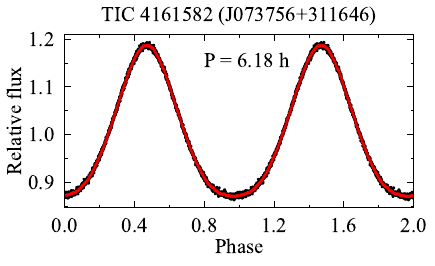}
    \end{minipage}
    \begin{minipage}{0.33\textwidth}
    \includegraphics[width=\textwidth]{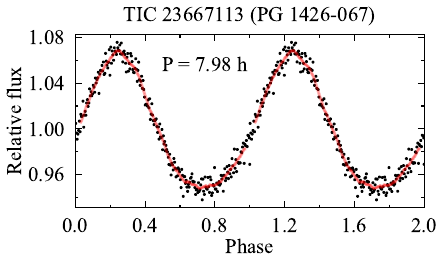}
    \end{minipage}
    \begin{minipage}{0.33\textwidth}
    \includegraphics[width=\textwidth]{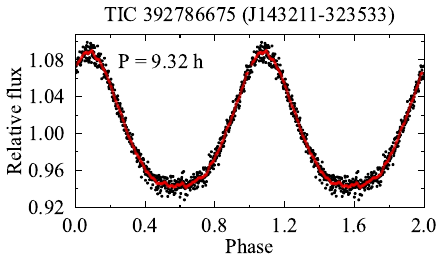}
    \end{minipage}
    \begin{minipage}{0.33\textwidth}
    \includegraphics[width=\textwidth]{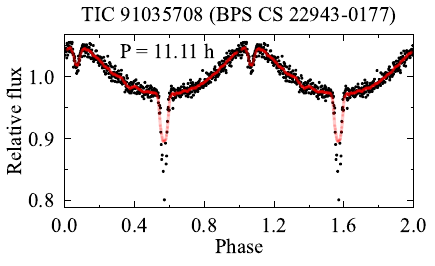}
    \end{minipage}
    \begin{minipage}{0.33\textwidth}
    \includegraphics[width=\textwidth]{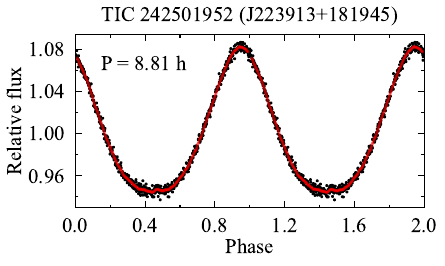}
    \end{minipage}
   \caption{Phase-folded light curves of 6 newly discovered reflection effect systems (with or without eclipse).}
    \label{Fig.2}
    \end{figure*}

The Lomb-Scargle periodograms \citep{Lomb.etal.1976,Scargle.etal.1982} of all light curves were calculated up to the Nyquist frequency. We phase-folded the light curves to the period determined by the periodogram or twice the period for ellipsoidal deformation systems. Subsequently, we visually inspected the periodograms and the phase-folded light curves for all targets to confirm any variability. Figure \ref{Fig.1} shows an example light curve and the periodogram of a reflection effect system that was derived through the custom script developed by \citet{Pelisoli.etal.2020}. A detailed introduction about the process of deriving phase-folded light curves can be found in \citet{Pelisoli.etal.2020}.

\subsection{The detection of reflection effect systems}

\begin{figure}
    {
    \centering
    \includegraphics[width=0.48\textwidth]{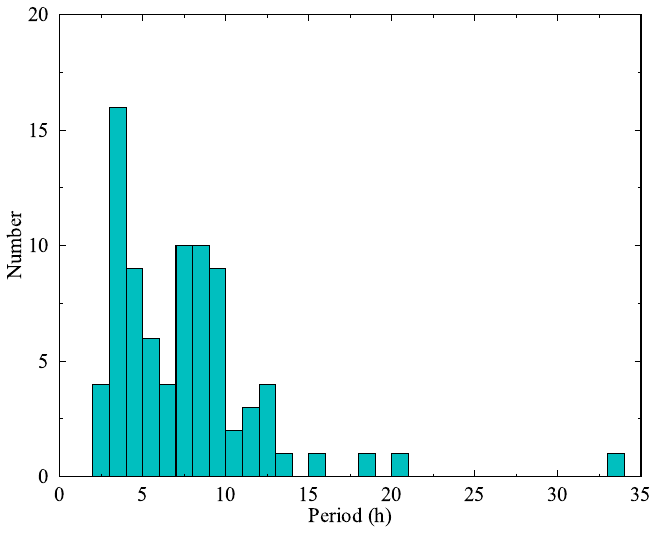}
      \caption{Period distribution of our detected reflection effect systems. 
              }
         \label{Fig.3}
         }
    \end{figure}

After conducting visual inspections of the light curves and the periodograms of 3,195 hot subdwarfs, we discovered a total of 82 reflection effect systems (with or without eclipse). Among them, six are newly discovered, while the others have been confirmed as reflection effect systems by \citet{Schaffenroth.etal.2019,Schaffenroth.etal.2022,Barlow.etal.2022,Sahoo.etal.2020,Krzesinski.etal.2022,Baran.etal.2021MNRAS.503.2157B,Baran.etal.2021,Kawka.etal.2015,Kupfer.etal.2015} and references therein. Figure \ref{Fig.2} shows the phase-folded light curves of our newly found reflection effect systems. In Figure \ref{Fig.3} we also present the period distribution of all our detected reflection effect systems. Most of them have periods less than half of one day. The period distribution we obtained has a similar trend to that discovered by \citet{Schaffenroth.etal.2022}.

    \begin{figure}
    {
    \centering
    \includegraphics[width=0.48\textwidth]{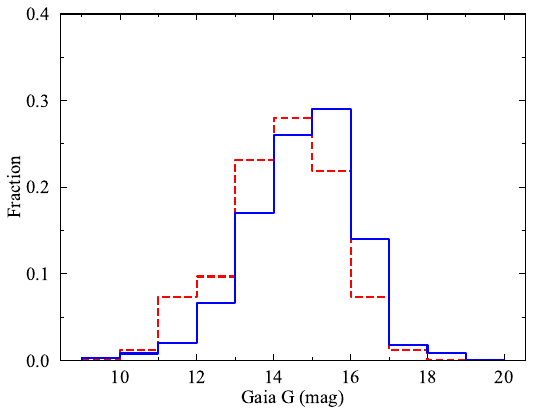}
      \caption{$G$aia $G$-band magnitude distribution of all our 3,195 hot subdwarfs with TESS observations, and a comparison with the distribution of the detected reflection effect systems. The solid blue and dashed red lines represent the $G$ magnitude distributions of all our samples and the reflection effect systems, respectively. 
              }
         \label{Fig.4}
         }
    \end{figure}

The brightness of stars can influence the shape and signal-to-noise ratio (S/N) of light curves and might impact the detection efficiency of reflection effect systems. To inspect whether our detection of reflection effect systems was influenced by their brightness, we present the $G$aia $G$-band magnitude distribution of all our samples in Figure \ref{Fig.4}, and make a comparison with the distribution of the reflection effect systems. Most of our hot subdwarfs have $G$ magnitudes brighter than 16 mag, and the $G$ magnitude distribution of the reflection effect systems exhibits a similar trend to the whole sample. The Kolmogorov-Smirnov (K-S) test of these two distributions also does not show any statistically significant difference ($ P_{\textup{KS}}$ = 0.86). Since most of the reflection effect systems were discovered with periods less than half of one day, the brightness seems not to obviously influence their detection efficiency in such short orbital periods. Meanwhile, for the reflection effect systems brighter than 15 mag even with a low inclination of only $10^{\circ}$, \citet{Schaffenroth.etal.2022} suggest that we can detect them up to two days by using the TESS light curves, as shown in the Figure 15 of \citet{Schaffenroth.etal.2022}. We expect to detect most of the reflection effect systems among our samples because the majority of them might have short periods less than half of one day and their light variations are obvious. Nevertheless, some of the reflection effect systems with relatively long orbital periods or low inclinations might be missed. Therefore, the fractions of reflection effect systems in different subclasses of hot subdwarfs obtained by us are just their lower limits, and we mainly focus on comparing the differences among different subclasses of hot subdwarfs. 

\subsection{Other types of newly found light variable hot subdwarfs}

\begin{figure*}
    {
    \centering
    \includegraphics[width=0.48\textwidth]{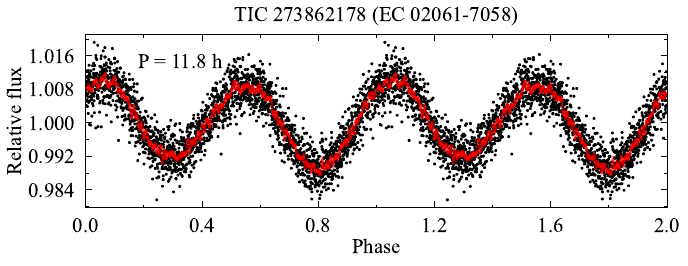}
    \includegraphics[width=0.48\textwidth]{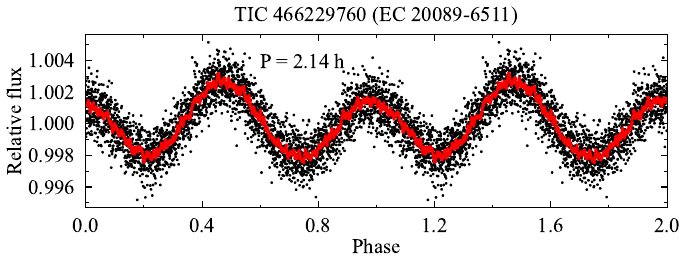}
      \caption{Phase-folded light curves of 2 newly discovered ellipsoidal deformation systems.
              }
         \label{Fig.5}
         }
    \end{figure*}

    \begin{figure*}
    {
    \centering
    \includegraphics[width=0.33\textwidth]{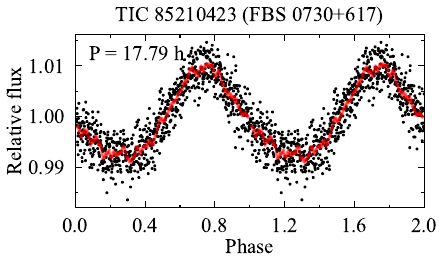}
     \includegraphics[width=0.33\textwidth]{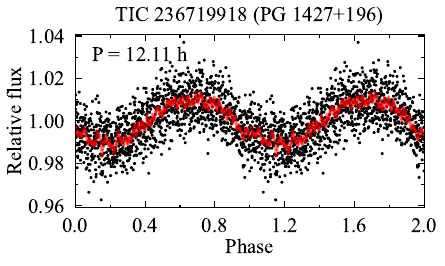}
    \includegraphics[width=0.33\textwidth]{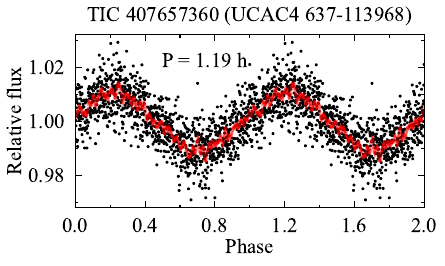}
      \caption{Phase-folded light curves of 3 newly discovered light-variable He-rich hot subdwarfs.
              }
         \label{Fig.6}
         }
    \end{figure*}

    \begin{table*}
 \caption{The main parameters of our newly discovered light-variable hot subdwarfs.}
 \label{Table.1}
 \centering
 \fontsize{8.5pt}{11pt}\selectfont
 \begin{tabular}{cccccccccc}
 \hline
 \hline
 \noalign{\smallskip}
    R.A. & Decl. & $T_{\rm {eff}}$ & log g & $\log n{\rm (He)}/n{\rm (H)}$ & Spclass & TIC & Type & Period & Likelihood \\ 
    (deg) & (deg) & (K) & (cm ${\rm s^{-2}}$) & ~ & ~ & (ID) & ~ & (h) & (\%)\\ 
    \noalign{\smallskip}
    \hline 
    \noalign{\smallskip}
    220.220106 & -3.147965 & 29490 $\pm$ 260 & 5.45 $\pm$ 0.03 & -2.90 $\pm$ 0.11 & sdB & 272266194 & Reflection & 8.06 & 99.9 \\
    114.48439 & 31.279597 & 31000 $\pm$ 290 & 5.49 $\pm$ 0.05 & -2.42 $\pm$ 0.09 & sdB & 4161582 & Reflection & 6.18 & 99.9 \\
    217.2137055 & -6.951050469 & ... &  ...  & ... & sdB & 23667113 & Reflection & 7.98 & 99.9 \\
    218.0489212 & -32.59297752 & ... &  ...  & ... & sdB & 392786675 & Reflection & 9.32 & 91.1 \\
    308.9570059 & -46.62934394 & ... &  ...  & ... & sdB & 91035708 & HW-Vir & 11.11 & 99.9 \\
    339.806741 & 18.32947834 & 28970 $\pm$ 390 & 5.44 $\pm$ 0.08 & -2.55 $\pm$ 0.21 &  sdB & 242501952 & Reflection & 8.81 & 99.9 \\
    31.7558425 & -70.74217757 & ... &  ...  & ... & sdB	& 273862178	& Ellipsoidal &	11.8 & 94.9 \\
    303.3811507 & -65.03518382 & ... &  ...  & ... & sdB & 466229760 & Ellipsoidal & 2.14 & 99.9 \\
    113.8625301 & 61.54535441 & 48400 $\pm$ 1410 & 6.05 $\pm$ 0.21 & 0.04 $\pm$ 0.56 & iHe-sdO & 85210423 & Variable & 17.79 & 99.9 \\
    217.3656669 & 19.36015037 & 50580 $\pm$ 1460 & 5.55 $\pm$ 0.16 & 0.29 $\pm$ 0.06 & iHe-sdO & 236719918 & Variable & 12.11 & 99.9 \\
    326.5013079 & 37.3554612 & 53123 $\pm$ 459 & 5.93 $\pm$ 0.06 & 1.20 $\pm$ 0.22 & eHe-sdO & 407657360 & Variable & 1.19 & 99.9 \\ \hline
    \noalign{\smallskip}
 \end{tabular}
 \end{table*}

    \begin{figure*}
    {
    \includegraphics[width=0.33\textwidth]{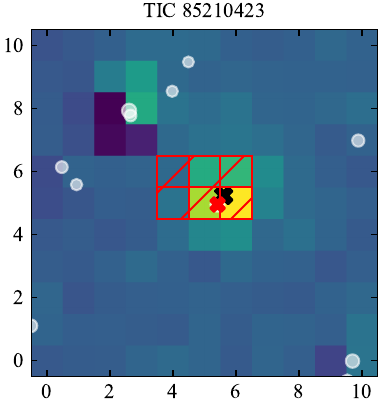}    
     \includegraphics[width=0.33\textwidth]{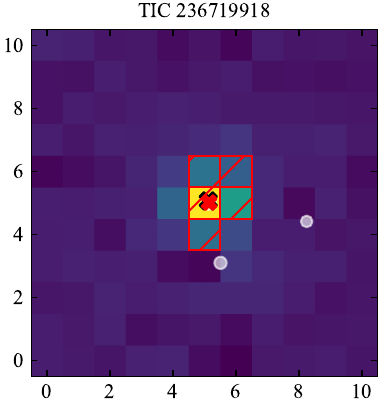}
     \includegraphics[width=0.33\textwidth]{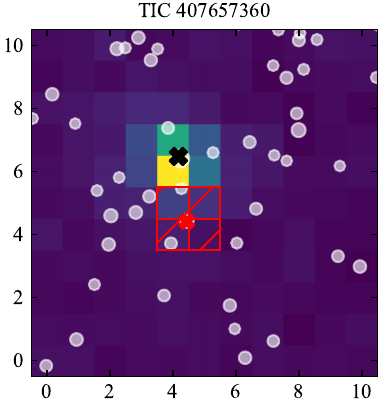}
      \caption{Example heat maps of three variable He-rich hot subdwarfs calculated through TESS-Localize. The target star is marked by a red cross, and the black cross represents the localized signal. The TESS aperture is marked with red borders, and the stars in the TPF are marked in white circles.
              }
         \label{Fig.7}
         }
    \end{figure*}

In the process of inspecting light curves, we also discovered 2 ellipsoidal deformation systems and 3 light-variable He-rich hot subdwarfs that were previously unknown. Their phase-folded light curves are plotted in Figure \ref{Fig.5} and Figure \ref{Fig.6}, respectively. It should be noted that we mainly focus on the reflection effect systems, and the discovery of the ellipsoidal deformation systems and light-variable He-rich hot subdwarfs is a by-product of our work. The information of our detected light-variable hot subdwarfs is shown in Table \ref{Table.A1} and the full table is available in electronic form. We also separately present the main parameters of all our newly discovered light-variable hot subdwarfs in Table \ref{Table.1}.

    \begin{figure*}
    {
    \centering
    \includegraphics[width=0.78\textwidth]{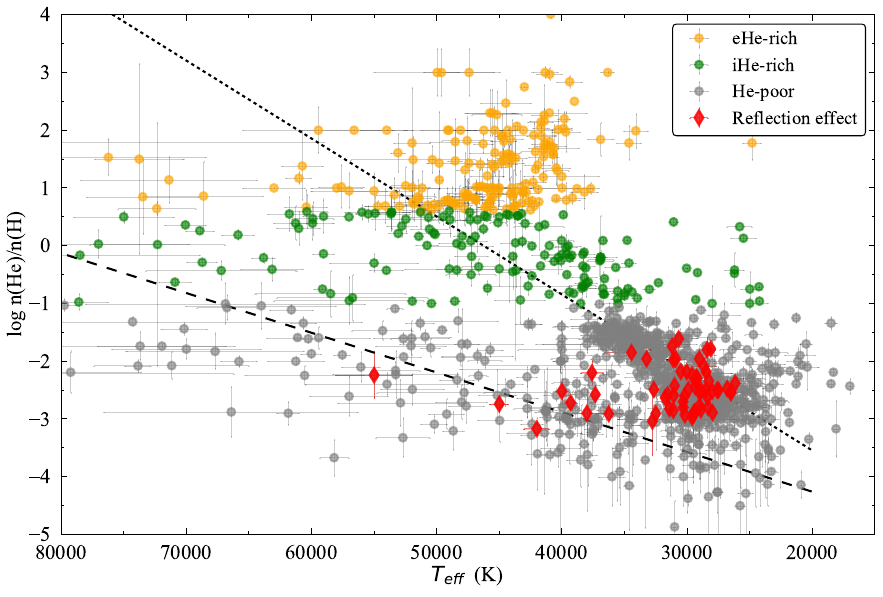}
      \caption{$T_{\rm eff}$ -- $\log n{\rm (He)}/n{\rm (H)}$ diagram of hot subdwarfs with reliable atmospheric parameters. The red diamonds represent the hot subdwarfs with the reflection effect. Hot subdwarfs that were not discovered with the reflection effect are labeled by circles, and the gray, green, and orange circles represent He-poor, iHe-rich, and eHe-rich hot subdwarfs, respectively. The dotted and dashed lines show the linear correlations between the $T_{\rm eff}$ and $\log n{\rm (He)}/n{\rm (H)}$ of hot subdwarfs fitted by \citet{Edelmann.etal.2003} and \citet{Peter.etal.2012}, respectively.
              }
         \label{Fig.8}
         }
    \end{figure*}

EC 02061-7058 was initially identified by \citet{Kilkenny.etal.2015} during the Edinburgh-Cape (EC) Blue Object survey and was classified as an sdB star. Its light curve shows variations with a period of 11.8 h. The differences in the minima of its light curve might be caused by the gravity darkening, while the differences in the maxima might be caused by the Doppler beaming effect. EC 20089-6511 was first classified as an sdB star by \citet{O'Donoghue.etal.2013}. Similar to EC 02061-7058, its light curve exhibits two distinct minima and maxima, but with a shorter period of 2.14 h and a much lower amplitude of variations.

Two iHe-rich hot subdwarfs and one eHe-rich hot subdwarf were discovered with quasi-sinusoidal variations in their light curves, and no obvious IR excess can be detected in their SEDs. FBS 0730+617 is an iHe-sdO star whose atmospheric parameters were first measured by \citet{Peter.etal.2012}. Its light curve shows a low amplitude of variations with a relatively long period of 17.79 h. PG 1427+196 was identified by \citet{Lei.etal.2019} as an iHe-sdO star. It shows long-period light variations of 12.11 h similar to FBS 0730+617. The reason for the light variations of these two iHe-sdO stars is still unknown. UCAC4 637-113968 is an eHe-sdO star which was identified by \citet{Lei.etal.2018}. It shows light variations with a short period of 1.19 h. However, UCAC4 637-113968 has a very low CROWDSAP value of 0.15, which might be contaminated by the nearby star. In order to further verify whether the variable signals originate from these He-rich hot subdwarfs, we used the TESS-Localize package\footnote{\url{https://github.com/Higgins00/TESS-Localize}} developed by \citet{Higgins.etal.2023} to determine the best-fit locations for the signal sources within the provided target pixel file (TPF) data and the measured frequencies from periodogram analysis, as recently shown in \citet{Krzesinski.etal.2025}. The software also calculates the value of relative-likelihood of each star’s location corresponding to the fit location. The calculated example heat maps of the amplitude fits for the three newly discovered photometric variable He-rich hot subdwarfs are shown in Figure \ref{Fig.7}. They show the best-fit location for the input frequencies as a black cross, and the location of target is marked with a red star. The positions of Gaia DR2 sources brighter than 18 mag within the frames are marked with white circles. We can see that the variable signals of TIC 85210423 and TIC 236719918 determined by TESS-Localize match well with the targets. However, the variable signals of TIC 407657360 are fitted to a contaminant star with a relative-likelihood of 99.9\%, indicating the variability does not originate from TIC 407657360. We also calculated the heat maps of the amplitude fits for our other newly discovered photometric variable hot subdwarfs by using TESS-Localize. The variability in all of them is fitted to the targets with the calculated relative-likelihood higher than 90\%, as shown in the last column of Table \ref{Table.1}, indicating the targets are the origin of the variability.

 \begin{figure*}
    {
    \centering
    \includegraphics[width=0.8\textwidth]{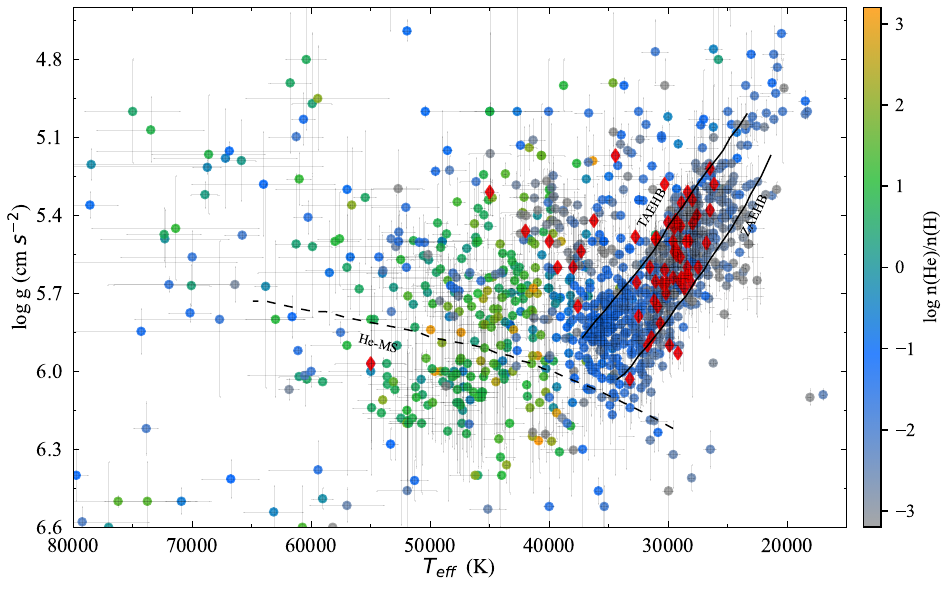}
      \caption{$T_{\rm eff}$ -- $\log g$ diagram of hot subdwarfs. Similar to Figure \ref{Fig.8}, the red diamonds represent hot subdwarfs with the reflection effect, and hot subdwarfs that were not detected with the reflection effect are labeled by circles. Furthermore, for these hot subdwarfs not showing the reflection effect, we also present their He-abundances with different color circles, as shown by the color bar. The zero-age EHB (ZAEHB) and terminal-age EHB (TAEHB) sequences with Z = 0.02 from \citet{Dorman.etal.1993} are presented as solid lines. The helium main sequence (He-MS) from \citet{Paczynski.etal.1971} is marked by the dashed line.
              }
         \label{Fig.9}
         }
    \end{figure*}

\section{The fraction of reflection effect systems in hot subdwarfs}\label{sect.3}

It is an important subject to investigate the fractions of reflection effect systems in different subclasses of hot subdwarfs. This can help us to estimate the fraction of hot subdwarfs with close low-mass M-type or brown dwarf companions and might provide new insights into the possible formation channels of different subclasses of hot subdwarfs. In this section, we divided our samples into different subclasses to count the fractions of reflection effect systems among them, and we studied the distribution features of the reflection effect systems in the $T_{\rm eff}$ -- $\log n{\rm (He)}/n{\rm (H)}$ diagram, the $T_{\rm eff}$ -- $\log g$ diagram, and the color-absolute magnitude diagram.

\subsection{Variability versus helium abundance}

To discuss in detail the fractions of reflection effect systems among different subclasses of hot subdwarfs, we only selected sources from our sample with atmospheric parameter measurements ($T_{\rm eff}$, $\log g$, and $\log n{\rm (He)}/n{\rm (H)}$). Furthermore, the atmospheric parameters of composite hot subdwarfs might not be accurate enough. Composite hot subdwarfs that were classified as sdO/B+F/G/K in \citet{Culpan.etal.2022} were removed. Then we used the Virtual Observatory SED Analyzer\footnote{\url{http://svo2.cab.inta-csic.es/svo/theory/vosa/}} \citep[VOSA,][]{Bayo.etal.2008} of the Spanish Virtual Observatory (SVO) to build the SEDs for our targets. Hot subdwarfs showing composite-SEDs were excluded. Finally, we obtained a sample of 1,225 stars with both reliable atmospheric parameters and TESS observations.

As shown in Figure \ref{Fig.8}, hot subdwarfs were classified into He-poor, iHe-rich, and eHe-rich according to their surface helium-abundances, and two distinct helium sequences can be seen. The first one (black dotted line) was discovered by \citet{Edelmann.etal.2003} and confirmed by \citet{Peter.etal.2012,Geier.2013A&A...557A.122G,Lei.etal.2019}, and so on. The helium-abundances of hot subdwarfs in this sequence exhibit a steeper increasing trend as the temperatures increase compared to the second sequence (black dashed line, discovered by \citet{Peter.etal.2012}), while the eHe-rich hot subdwarfs do not follow the relationship. We found that the reflection effect systems are scattered on both the helium sequences but with a main concentration on the first one, and they exhibit highly uneven distributions among hot subdwarfs with different helium abundances. About 7\% of our He-poor hot subdwarfs were detected with the reflection effect, while none of the iHe-rich and eHe-rich hot subdwarfs were found with the obvious signal of the reflection effect. In the upper panel of Table \ref{Table.2} we present the fractions of reflection effect systems detected among hot subdwarfs within different helium abundance ranges. The uncertainties in the fractions were derived based on the assumption of a binomial distribution, while the uncertainties for the iHe-rich and eHe-rich hot subdwarfs without detectable reflection effect were estimated by using the beta distribution to calculate the 68\% confidence upper limit. 

Except for their different atmospheric chemical compositions, He-rich and He-poor hot subdwarfs also exhibit distinct fractions of reflection effect systems and binarity, which indicate that they might originate from different formation channels. Most of the He-rich hot subdwarfs are believed to be single stars and they are thought to form via the merger channels such as the double He-WDs merger \citep[e.g.,][]{Zhang.etal.2012,Geier.etal.2022}. The special formation mechanism for He-rich hot subdwarfs can well explain their surface helium enrichment, as well as their low fraction of reflection effect systems and low binary fraction. On the other hand, the higher fraction of reflection effect systems and binary fraction for He-poor hot subdwarfs indicate the CE ejection channel might play an important role in their formation. 

\begin{table}
    \fontsize{9.3pt}{10.3pt}\selectfont
    \caption[]{The fractions of reflection effect systems among different subclasses of hot subdwarfs.
    }
    \label{Table.2}
    \begin{tabular}{cccccccc}
    \hline
    \noalign{\smallskip}
    Subclass      & Total &   Reflection   &   Fraction (\%)  &   \\ \hline
    \noalign{\smallskip}
    eHe-rich      & 178 & 0  & $0^{+0.5}_{-0}$ &  \smallskip\\
    iHe-rich    &  153 & 0  & $0^{+0.6}_{-0}$ &   \\ 
    He-poor    & 894  & 61 & 6.8 $\pm$ 1.6 &   \\  
    \noalign{\smallskip} \hline
     EHBa  & 80  & 2 & 2.5 $\pm$ 1.7 &   \\
     EHBb  & 335  & 42 & 12.5 $\pm$ 1.8 & \\
     EHBc  & 182  & 3 & 1.6 $\pm$ 0.9 &   \\
     bEHB  & 126  & 3 & 2.4 $\pm$ 1.4 &   \\
     postEHBa  & 108  & 10 & 9.3 $\pm$ 2.8 &  \\
     postEHBb  & 63  & 1 & 1.6 $\pm$ 1.6 &  \\
    \noalign{\smallskip}  \hline  
    \end{tabular}
    \end{table} 

Although the majority of He-rich hot subdwarfs are found to be single stars, several special He-rich hot subdwarfs were also detected in short-period binaries that formed through very peculiar channels \citep{Ahmad.etal.2004,Sener.etal.2014,Ratzloff.etal.2020,Reindl.etal.2020,Kupfer.etal.2020b,Kupfer.etal.2020a,Snowdon.etal.2023,Luochangqing.etal.2024}, but none of them were found with low-mass M-type or BD companions. In a recent study that searched for short-period He-rich hot subdwarf binaries, \citet{Snowdon.etal.2025} also did not detect any obvious signal of the reflection effect among the 29 iHe-rich hot subdwarfs and 124 eHe-rich hot subdwarfs identified through the Southern African Large Telescope (SALT) survey. Up to now, only \citet{Barlow.etal.2022} have discovered an eHe-rich sdO star (J278.2669+46.6181) showing a strong reflection effect with a period of about 1.7 h. Short-period He-rich hot subdwarf binaries with low-mass M-type or BD companions are extremely rare. They might have a very low formation rate, or they are few in number and might have been missed when inspecting their light curves or RV variations.

\subsection{Variability along the different regions of the extreme horizontal branch}

\begin{figure}
    {
    \centering
    \includegraphics[width=0.48\textwidth]{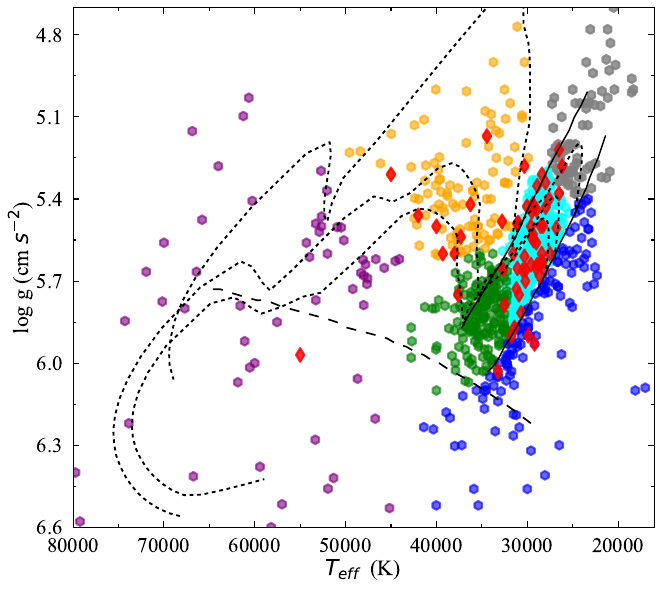}
      \caption{$T_{\rm eff}$ -- $\log g$ diagram of different subclasses of He-poor hot subdwarfs along the different regions of the EHB. Different subclasses of hot subdwarfs are marked by different color pentagons, for which the EHBa, EHBb, EHBc, bEHB, postEHBa, and postEHBb stars are labeled by the gray, cyan, green, blue, orange, and purple pentagons, respectively. The detected reflection effect systems are highlighted by the red diamonds. Three evolutionary tracks of hot subdwarfs with a He-core mass of 0.48 $M_\odot$ and different hydrogen envelope masses calculated by \citet{Dorman.etal.1993} are marked by the dotted lines.
              }
         \label{Fig.10}
         }
    \end{figure}

In Figure \ref{Fig.9} we present the distributions of hot subdwarfs along the different regions of the EHB. Their helium abundances and the detected reflection effect systems are also shown. We found that the reflection effect systems are not uniformly distributed in the different regions of the EHB. Most of them are located in the region where He-poor hot subdwarfs are grouped, and they show a similar distribution trend with the significant RV-variable He-poor hot subdwarfs as discovered by \citet{Geier.etal.2022} and \citet{Ruijie.etal.2024}. To discuss this feature in detail, we divided our He-poor hot subdwarfs into EHBa, EHBb, EHBc, bEHBa, postEHBa, and postEHBb stars according to their locations on the EHB and the RV-variability characteristics of different subclasses, as shown in Figure \ref{Fig.10}. Different fractions of reflection effect systems were detected in the six subclasses of hot subdwarfs, as presented in the bottom panel of Table \ref{Table.2}.

The EHBa stars are located on the top of the EHB. Most of them have temperatures cooler than about 25,000 K, and only about 3\% of them were detected with the reflection effect. This number is smaller than the fraction of about 13\% for EHBb stars with $T_{\rm eff}$ $\sim$ 25,000 -- 33,000 K and $\log g$ $\sim$ 5.3 -- 5.9. The EHBc stars are concentrated on the bottom of the EHB (with $T_{\rm eff}$ $\sim$ 33,000 -- 40,000 K and $\log g$ $\sim$ 5.7 -- 6.0). The fraction of reflection effect systems detected in them is about 2\%. Both the lower fractions of reflection effect systems for EHBa and EHBc stars are consistent with their lower detected RV-variability fractions, which indicates that EHBa and EHBc stars might have a lower short-period binary fraction and different formation channels to EHBb stars. Specifically, EHBa stars might contain a new subpopulation of long-period binaries with late-type MS or compact companions as speculated by \citet{Geier.etal.2022} or they might have a higher fraction of single stars that formed through the merger channels. For EHBc stars, most of them have higher helium abundances than EHBa and EHBb stars as shown in Figure \ref{Fig.9}, and their higher helium abundances are consistent with the prediction of \citet{Miller.etal.2008} based on diffusion processes of He-rich hot subdwarfs. The EHBc stars might have a higher fraction of single stars that evolved from He-rich hot subdwarfs \citep{Geier.etal.2022,Miller.etal.2008}. Since EHBb stars exhibit a higher fraction of reflection effect systems and RV-variable stars than EHBa and EHBc stars, the CE ejection channel might play a more important role in their formation.

Hot subdwarfs above the EHB were mainly classified as postEHBa and postEHBb stars. The lower temperature group with $T_{\rm eff}$ $\sim$ 35,000 -- 45,000 K was labeled as postEHBa, while the hotter group with $T_{\rm eff}$ $\sim$ 45,000 -- 75,000 K was labeled as postEHBb. We found that these two groups of stars exhibit different fractions of reflection effect systems. About 9\% of postEHBa stars were detected with the reflection effect, which is consistent with that fraction for EHBb stars within the uncertainties. This supports the conclusion derived through their RV-variability that postEHBa stars might have an evolutionary connection to EHBb stars as shown by the evolutionary tracks in Figure \ref{Fig.10}. However, postEHBb stars exhibit a lower fraction of reflection effect systems of about 2\%, and their RV-variability fraction is also lower than that of EHBb and postEHBa stars. We speculate that some of the postEHBb stars might not be connected to the subsequent evolution of EHBb and postEHBa stars, and they seem to be single stars which might evolve from He-sdO stars by diffusion processes or originate from the subsequent evolution of EHBc stars.

\begin{figure}
    {
    \centering
    \includegraphics[width=0.48\textwidth]{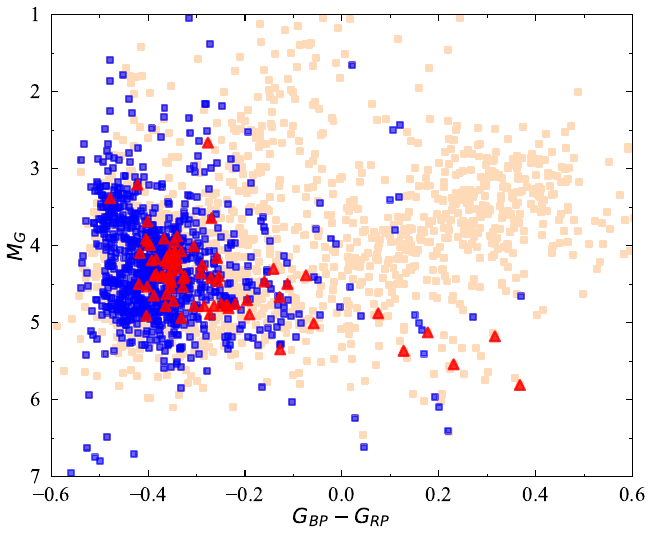}
      \caption{Color-absolute magnitude diagram of our samples with TESS observations. Hot subdwarfs with reliable atmospheric parameters are marked by blue squares, while those without reliable atmospheric parameters are labeled by light brown squares. The reflection effect systems are labeled by red triangles.
              }
         \label{Fig.11}
         }
    \end{figure}

    \begin{figure*}
    {
    \centering
    \includegraphics[width=0.8\textwidth]{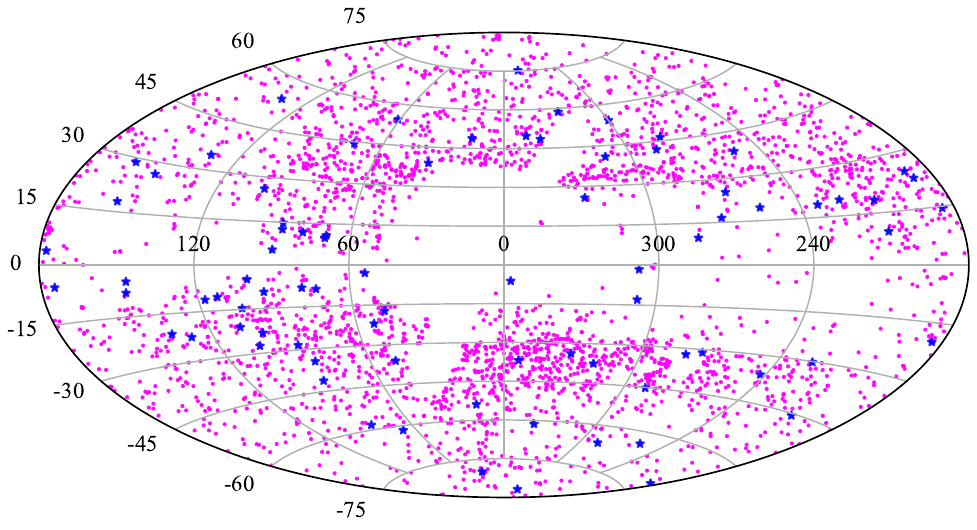}
      \caption{Positions of hot subdwarfs with TESS observations on the sky (in Galactic coordinates). The blue stars represent hot subdwarfs that were detected with the reflection effect, while hot subdwarfs that were not discovered with the reflection effect are labeled by magenta circles.
              }
         \label{Fig.12}
         }
    \end{figure*}

    \begin{figure}
    {
    \centering
    \includegraphics[width=0.48\textwidth]{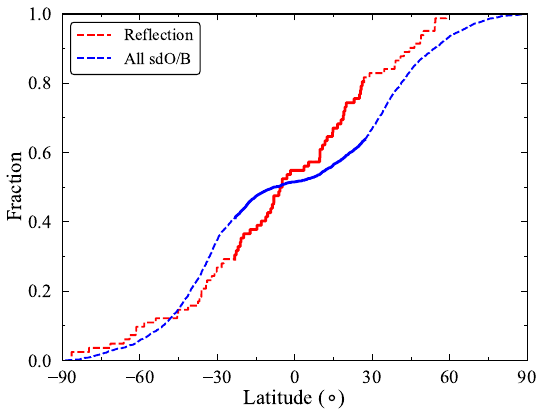}
      \caption{Cumulative distributions at latitude for hot subdwarfs that were detected with the reflection effect (red dashed line) or not (blue dashed line). In order to compare the differences between these two distributions near the Galactic Plane, we highlight the cumulative distributions of the reflection effect systems and the whole hot subdwarf sample within $-23^{\circ} < b < 27^{\circ}$ by the solid red line and the solid blue line.
              }
         \label{Fig.13}
         }
    \end{figure}

The bEHB stars are composed of hot subdwarfs below the canonical ZAEHB. A lower fraction of reflection effect systems of about 2\% were also detected in them. However, different from EHBa, EHBc, and postEHBb stars with both lower fractions of reflection effect systems and lower RV-variability fractions, bEHB stars exhibit a high RV-variability fraction and a higher fraction of stars with large RV-amplitudes than other subclasses of hot subdwarfs. \citet{Geier.etal.2022} suggest that hot subdwarfs below the EHB might be composed of less massive EHB stars that evolved from the intermediate-mass stars with ignited nondegenerate helium cores or progenitors of He-WDs (pre-He-WDs) with non-core helium burning. Compared to the canonical short-period hot subdwarfs that formed through the CE ejection at the tip of the red giant branch (RGB), both the less massive EHB stars and pre-He-WDs require more massive companions to eject the more tightly bound envelopes of their progenitors, and the companions probably have a deeper spiral-in \citep{Geier.etal.2022}. Therefore, they might have shorter orbital periods and higher RV-amplitudes. The lower fraction of reflection effect systems for bEHB stars indicates that they might have a low fraction of stars with M-type or brown dwarf companions. Another possibility for the companions of bEHB stars is the WD, and hot subdwarfs with WD companions generally exhibit higher RV-amplitudes than those with M-type or brown dwarf companions, as shown in the Figure 18 of \citet{Schaffenroth.etal.2022}. The lower fraction of reflection effect systems and the higher fraction of stars with large RV-amplitudes detected in bEHB stars might be because they have a higher fraction of stars with close WD companions.

\subsection{Distribution in the color-absolute magnitude diagram}

In Figure \ref{Fig.11} we present the distributions of hot subdwarfs with parallax precision better than 20\% in the color-absolute magnitude diagram. Hot subdwarfs with atmospheric parameters and the reflection effect systems are highlighted as blue squares and red triangles, respectively. The reflection effect systems are mainly concentrated at $-0.45< G_{\textup{BP}}-G_{\textup{RP}}< -0.1$ and $3.5< M_{\textup{G}}< 5.0$, and this region is mainly occupied by sdB stars. The hotter sdO and He-rich hot subdwarfs mainly group at the bluer end of the color-absolute magnitude diagram with $ G_{\textup{BP}}-G_{\textup{RP}}<-0.45$, but very few hot subdwarfs with the reflection effect are located at that region. It might be because the hotter sdO and He-rich hot subdwarfs have a low fraction of stars with close low-mass M-type or brown dwarf companions. The composite hot subdwarfs are mainly located at $0< G_{\textup{BP}}-G_{\textup{RP}}< 0.5$ and $2< M_{\textup{G}}< 4.5$, while no reflection effect systems are found there. The reason might be that most of the composite hot subdwarfs are long-period binaries.

\section{Positions of reflection effect systems on the sky}\label{sect.4}

As described in the introduction, hot subdwarfs close to the Galactic Plane might have a higher fraction of stars with the reflection effect than those at higher latitudes. To confirm the speculation, we present the distributions on the sky for all our samples that show the reflection effect or not, as seen in Figure \ref{Fig.12}. Among them, 716 hot subdwarfs are scattered within $-25^{\circ} < b < 25^{\circ}$, and 2479 objects are located at $\vert b \vert > 25^{\circ}$. A fraction of 5.3\% for hot subdwarfs in the lower latitude group were detected with the reflection effect, while a smaller value of 1.8\% for that fraction was found in hot subdwarfs at $\vert b \vert > 25^{\circ}$. To further verify whether the distributions near the Galactic Plane between the reflection effect systems and all the hot subdwarfs are statistically significant, we present the cumulative distributions of these two types of samples in terms of latitude as shown in Figure \ref{Fig.13}. We found that the cumulative curve for the detected reflection effect systems at $-23^{\circ} < b < 27^{\circ}$ exhibits a more rapidly increasing trend and a larger numerical change compared to the entire sample in the same latitude range. The KS-test between these two distributions ($ P_{\textup{KS}}$ < $10^{-10}$) also shows a significant difference. The above discoveries all indicate that hot subdwarfs near the Galactic Plane might have a higher fraction of stars with the reflection effect than the objects at higher latitudes. 
 
The mean age of the progenitors of hot subdwarfs was proposed by \citet{Moni-Bidin.etal.2008} as an important parameter that might influence the short-period binary fraction of hot subdwarfs, for which the younger populations are more likely to produce hot subdwarfs with higher short-period binary fraction and the older populations tend to form a lower short-period binary fraction. Subsequently, by conducting the detailed BPS simulations for the formations of hot subdwarfs through different channels, \citet{Han.etal.2008} found that the younger populations have a higher birth rate for the hot subdwarfs formed through the first and second CE ejection channels than the older populations. On the other hand, it is generally believed that stars around the Galactic Plane belong to the younger populations, while stars at higher latitudes are more likely to be the older populations. The higher fraction of reflection effect systems detected in hot subdwarfs close to the Galactic Plane than the objects at higher latitudes might imply that hot subdwarfs produced by the younger populations have a higher fraction of stars with close low-mass M-type or BD companions than those produced by the older populations. This might be important evidence for the prediction of \citet{Moni-Bidin.etal.2008} and \citet{Han.etal.2008}. Nevertheless, because of the erroneous astrometry and the dense environment for stars near the Galactic Plane, the number of hot subdwarfs that were confirmed in this region is not large enough. Here, we only conducted a limited discussion on the fractions of reflection effect systems detected among hot subdwarfs in different regions of the Galaxy. More accurate measurements and observations focused on the Galactic plane are needed to further verify the discovery.

\section{Summary and conclusions}\label{sect.5}

In this paper we have inspected the light curves of 3195 hot subdwarfs to estimate the fractions of reflection effect systems among different subclasses of hot subdwarfs. We have discovered a total of 82 reflection effect systems, and 6 of them are newly discovered. As a by-product of our research, we also discovered 2 ellipsoidal deformation systems and 3 light-variable He-rich hot subdwarfs that were previously unknown. Subsequently, we divided our samples into different subclasses according to their surface helium abundances and their locations along the different regions of the EHB to explore the fractions of reflection effect systems among these subclasses. Meanwhile, a discussion about the possible formation channels of different subclasses of hot subdwarfs was presented by combining their RV-variability characteristics. We also explored the distribution characteristics of the reflection effect systems in the Galaxy. Here we summarized the basic results as follows.

1. None of the iHe-rich or eHe-rich hot subdwarfs were discovered to have obvious characteristics of the reflection effect. This supports that most of He-rich hot subdwarfs might be single stars formed through the merger channels. 

2. He-poor hot subdwarfs along the different regions of the EHB were detected with distinct fractions of reflection effect systems. Only about 3\% of hot subdwarfs located on the top of EHB and about 2\% of hot subdwarfs located on the bottom of the EHB show the reflection effect. The two values are lower than that fraction of about 13\% for hot subdwarfs on the EHB with $T_{\rm eff}$ $\sim$ 25,000 -- 33,000 K and $\log g$ $\sim$ 5.3 -- 5.9. Hot subdwarfs located above the EHB with $T_{\rm eff}$ $\sim$ 45,000 -- 75,000 K were also detected with a low fraction of reflection effect systems of about 2\%. The results here are consistent with that derived through their RV-variability. These three subclasses of hot subdwarfs with lower fractions of reflection effect systems might have a higher fraction of single stars. Hot subdwarfs on the top of the EHB might also contain a subclass of stars with long-period cool or compact companions.

3. The fraction of reflection effect systems for He-poor hot subdwarfs located above the EHB with $T_{\rm eff}$ $\sim$ 35,000 -- 45,000 K is close to that fraction detected in hot subdwarfs on the EHB with $T_{\rm eff}$ $\sim$ 25,000 -- 33,000 K. These two subclasses of
hot subdwarfs might have an evolutionary connection.

4. Only about 2\% of the hot subdwarfs below the EHB were detected with the reflection effect. The majority of hot subdwarfs below the EHB might be low-mass hot subdwarfs or progenitors of He-WDs, and they might have a higher fraction of stars with close WD companions.

5. The reflection effect systems are not uniformly distributed in the Galaxy. About 5\% of hot subdwarfs within $-25^{\circ} < b < 25^{\circ}$ were detected with the reflection effect, while that fraction for hot subdwarfs located at $\vert b \vert > 25$ is only about 2\%. Hot subdwarfs close to the Galactic Plane might have a higher fraction of stars with close low-mass M-type or BD companions than those at higher latitudes. 

\section*{Acknowledgements}

We thank the anonymous referee for his/her valuable suggestions and comments which improved this work greatly. This work is supported by the National Natural Science Foundation of China (Nos. 12288102 and 12333008) and National Key R$\&$D Program of China (No. 2021YFA1600403). X.M. acknowledges support from Yunnan Fundamental Research Projects (Nos. 202401BC070007 and 202201BC070003), International Centre of Supernovae, Yunnan Key Laboratory (No. 202302AN360001), the Yunnan Revitalization Talent Support Program Science $\&$ Technology Champion Project (NO. 202305AB350003), and the science research grants from the China Manned Space Program with grant no. CMS-CSST-2025-A13. 

\section*{Data Availability}

The TESS light curves used in this study can be accessed from the MAST portal: \url{https://mast.stsci.edu}. A table with the information of all our detected photometric variable hot subdwarfs will be provided online and also available in electronic form at the CDS via anonymous ftp to \url{cdsarc.u-strasbg.fr} (\url{130.79.128.5}).



\bibliographystyle{mnras}
\bibliography{reference.bib} 
\appendix\section{Information about supplementary material}
\begin{table*}
\caption{The main parameters of our detected light-variable hot subdwarfs. The full table is available as supplementary material.}
\label{Table.A1}
\centering
\fontsize{8.5pt}{11pt}\selectfont
\begin{tabular}{cccccccccc}
\hline
\hline
\noalign{\smallskip}
    R.A. & Decl. & $T_{\rm {eff}}$ & log g & $\log n{\rm (He)}/n{\rm (H)}$ & Spclass & TIC & Type & Period & Crowdsap \\ 
    (deg) & (deg) & (K) & (cm ${\rm s^{-2}}$) & ~ & ~ & (ID) & ~ & (h) & ~ \\ 
    \noalign{\smallskip}
    \hline 
    \noalign{\smallskip}
202.973147	&	15.688212	&	29370	$\pm$	250	&	5.53	$\pm$	0.04	&	-2.87	$\pm$	0.28	&	sdB	&	95526898	&	Reflection	&	5.99	&	0.84	\\
312.413176	&	30.081818	&	37330	$\pm$	170	&	5.54	$\pm$	0.06	&	-2.58	$\pm$	0.12	&	sdOB	&	230775376	&	Reflection	&	10.31	& 0.65 \\
220.220106	&	-3.147965	&	29490	$\pm$	260	&	5.45	$\pm$	0.03	&	-2.90	$\pm$	0.11	&	sdB	&	272266194	&	Reflection	&	8.06	&	0.87	\\
346.401764	&	34.698363	&	28260	$\pm$	160	&	5.50	$\pm$	0.02	&	-2.34	$\pm$	0.03	&	sdB	&	369581468	&	Reflection	&	4.77	&	0.93\\
204.70065	&	-2.030358	&	32500	$\pm$	50	&	5.79	$\pm$	0.01	&	-2.91	$\pm$	0.08	&	sdB	&	175402069	&	HW-Vir	&	2.42	&	0.96\\
233.456025	&	37.991129	&	29300	$\pm$	120	&	5.55	$\pm$	0.01	&	-2.34	$\pm$	0.03	&	sdB	&	148785530	&	HW-Vir	&	3.88	&	0.97\\
117.23262	&	30.713059	&	31080	$\pm$	90	&	5.65	$\pm$	0.02	&	-1.98	$\pm$	0.04	&	sdB	&	4491131	&	Reflection	&	5.53	&	0.85\\
114.48439	&	31.279597	&	31000	$\pm$	290	&	5.49	$\pm$	0.05	&	-2.42	$\pm$	0.09	&	sdB	&	4161582	&	Reflection	&	6.18	&	0.43\\
71.237089	&	14.363909	&	32660	$\pm$	320	&	5.66	$\pm$	0.05	&	-2.49	$\pm$	0.13	&	sdB	&	436579904	&	Reflection	&	9.55	& 0.61\\
291.539417	&	37.335611	&	30670	$\pm$	160	&	5.82	$\pm$	0.05	&	-1.61	$\pm$	0.04	&	sdB	&	137306463	&	Reflection	&	7.02	& 0.62\\
 \hline
    
    \noalign{\smallskip}
\end{tabular}
\end{table*}



\label{lastpage}
\end{document}